# Out-Of-Place debugging: a debugging architecture to reduce debugging interference

## Matteo Marra[a], Guillermo Polito[b], and Elisa Gonzalez Boix[a]

a   Vrije Universiteit Brussel, Brussels, Belgium
b   Univ. Lille, CNRS, Centrale Lille, Inria, UMR 9189 - CRIStAL - Centre de Recherche en Informatique Signal et Automatique de Lille, F-59000 Lille, France

**Abstract**    **Context** Recent studies show that developers spend most of their programming time testing, verifying and debugging software. As applications become more and more complex, developers demand more advanced debugging support to ease the software development process.

**Inquiry** Since the 70's many debugging solutions have been introduced. Amongst them, online debuggers provide good insight on the conditions that led to a bug, allowing inspection and interaction with the variables of the program. However, most of the online debugging solutions introduce *debugging interference* to the execution of the program, i.e. pauses, latency, and evaluation of code containing side-effects.

**Approach** This paper investigates a novel debugging technique called *out-of-place* debugging. The goal is to minimize the debugging interference characteristic of online debugging while allowing online remote capabilities. An *out-of-place* debugger transfers the program execution and application state from the debugged application to the debugger application, each running in a different process.

**Knowledge** On the one hand, *out-of-place* debugging allows developers to debug applications remotely, overcoming the need of physical access to the machine where the debugged application is running. On the other hand, debugging happens locally on the remote machine avoiding latency. That makes it suitable to be deployed on a distributed system and handle the debugging of several processes running in parallel.

**Grounding** We implemented a concrete out-of-place debugger for the Pharo Smalltalk programming language. We show that our approach is practical by running several benchmarks, comparing our approach with a classic remote online debugger. We show that our prototype debugger outperforms a traditional remote debugger by 1000 times in several scenarios. Moreover, we show that the presence of our debugger does not impact the overall performance of an application.

**Importance** This work combines remote debugging with the debugging experience of a local online debugger. Out-of-place debugging is the first online debugging technique that can minimize debugging interference while debugging a remote application. Yet, it still keeps the benefits of online debugging (e.g., step-by-step execution). This makes the technique suitable for modern applications which are increasingly parallel, distributed and reactive to streams of data from various sources like sensors, UI, network, etc.

## The Art, Science, and Engineering of Programming

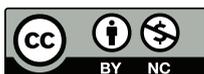

 **Introduction**

Debugging is one of the main activities in the development of modern applications [17]. A recent study conducted by Cambridge showed that most of the programming time is nowadays spent on testing, verifying and debugging software [4]. As software gets more complex, bugs get more sophisticated. Many difficulties raise when debugging parallel programs [30]. For instance, bugs may be caused by a particular interaction between many parallel components, or seen sporadically in concurrent processes. Understanding, replicating and solving such bugs is a challenging task that might take quite a lot of time and resources.

Over the years several techniques were proposed to aid developers in the arduous task of debugging (section 2). In particular, there exist two well-known families of debuggers: online and offline debuggers [25, 30]. Online debuggers control the execution of an application at the moment of a failure. They allow developers to interact smoothly with the target application, offering stepping and breakpoint operations that give immediate feedback to the developer. On the other hand, offline debuggers (or post-mortem debuggers) try to reconstruct the context of a bug once the process that failed is already finished. Such solutions analyse or replay log files, code dumps and/or execution traces to help the developer discover the source of the problem. Reproducing a bug with these techniques can be tedious and time-consuming, especially because many debugging cycles are required before the error happens again.

Online solutions provide good insight into the conditions that led to a bug since they allow developers to inspect the state of the application, and to interact with it through the evaluation of expressions. However, online debuggers introduce *debugging interference* (also known as the *probe effect* [25])[1] due to the communication of processes and the performance penalties introduced by the debugging infrastructure. Such interference alters the behaviour of the application and makes a bug more difficult to reproduce during debugging.[2]

**Contribution**   In this paper we present a novel debugging technique called *out-of-place* debugging (section 3). The goal is to combine the benefits of local and remote debugging, providing the latency of a local debugger while allowing developers to debug remote applications and scoping eventual side effects. When the debugged application is paused in a breakpoint or an exception, the execution stack and the application state are entirely transferred to the (remote) out-of-place debugger. The developer then proceeds to debug the application locally, with an entire copy of the program state.

In this paper, we show the benefits of *out-of-place* debugging in terms of reduced latency and scoping of side effects, in different debugging scenarios. We also present our implementation of IDRA, a prototype of *out-of-place* debugger in Pharo Smalltalk.

---

[1] Debugging interference is also known in computer science as the *the observer effect*. This states that an observer modifies the observed object while observing it.

[2] A bug that disappears during debugging it is also known as a *heisenbug*.

We show that the presence of our debugger does not impact application performance during normal execution. We also show that our prototype debugger performs better than a traditional remote debugger in some scenarios and can improve the debugging experience.

## 2 Motivation

To show the kind of problems that arise while debugging modern applications, we introduce two debugging scenarios: (1) a distributed and parallel application that analyzes tweets and (2) a cyber physical system (CPS) that is remotely deployed and listening to data coming from a sensor.

### 2.1 Debugging Scenario 1: Twitter application

Let us consider a distributed application that continuously analyzes tweets coming from a stream. For the purposes of this presentation, the relevant code of such an application is illustrated in listing 1. There are two methods: analyzeTweets that consumes a stream of tweets and processTweet: that creates a dictionary mapping each word occurring in the tweet with the number of occurrences in the tweet.

**■ Listing 1**   A simple tweet consuming application

```
1  TwitterApplication>>analyzeTweets
2    | results |
3    results := OrderedCollection new.
4    [ twitterStream hasNext ] whileTrue: [ | tweet |
5        tweet := twitterStream next.
6        results nextPut: (self processTweet: tweet).
7      ]
8    ↑self mergeDictionaries: results.
9
10 TwitterApplication>>processTweet: aTweet
11   | wordCount tweetObject text words |
12   tweetObject := self parseTweet: aTweet.
13   text := tweetObject text.
14   words := text findTokens.
15   wordCount := Dictionary new.
16   words do: [ :word |
17     (wordCount includesKey: word)
18       ifTrue: [ wordCount at: word put: 1+ (result at: word) ]
19       ifFalse: [ wordCount at: word put: 1 ]. ]
20   ↑result
```

At first glance, the code looks correct and on the developer's local machine the code runs without problems. Now consider the same application deployed on a cluster, executed in parallel. In this scenario, the distributed setup makes it even more difficult to track bugs.

Concretely in this case, if the application is left running for hours or even days, depending on the number of tweets in the stream, an out-of-memory error happens. Our twitter stream is optimized with a read buffer to read and allocate memory in chunks of more optimal sizes. The developer that designed the buffering optimization accidentally introduced a bug: the buffered objects are not correctly released, generating a memory leak.

Understanding the root cause of the problem is not an easy task. First, it is not easy to reproduce the conditions that make the bug appear. Even reproducing it in a production environment may take quite a long time for the error to appear. Second, the developer needs to perform several debugging cycles before finding the actual cause of the bug.

### 2.2 Debugging scenario 2: A temperature monitoring application

Another common scenario nowadays is in the domain of cyber physical systems (CPS). CPSs are applications that collect and process data from physical sources such as sensors. These systems are often deployed physically close to their data sources. Our scenario involves a CPS that monitors the temperature of a room. This monitoring system is made of a small computer, (concretely, a *Raspberry pi* [9], referred to as the device) connected to a temperature sensor and an LCD screen. We deploy the device in a room that we are interested in monitoring.[3] The sensor probes the room's temperature and displays the result on the LCD screen. This device is connected to the network via WiFi or ethernet and is configured to send alarms to the end-user if the temperature of the room exceeds a given level (e.g., in a food storage room). The internet connection is bi-directional: the device can also receive updates such as user configuration and firmware updates.

When testing our application, the device works fine the majority of the time. However, from time to time false alarms are sent to the user. Restarting the device solves the problem temporarily: after an undetermined period of time the bug reappears. Reproducing the bug is not easy because we cannot predict the timing of the bug. In addition, in production mode the temperature monitor works remotely, so when there is a problem we cannot know for sure what it is happening. Hence, reproducing the exact conditions under which the bug happens is complicated.

### 2.3 Online and Offline Debugging

In this section, we discuss existing debugging techniques and how they apply to modern applications as the ones described in the previous sections. Following Pacheco [30] and the survey from McDowell and Helmbold [25], we categorize debugging techniques in two big families: *Offline Debugging* and *Online Debugging*.

---

[3] More information about the CPS application and its deployment can be found in [24].

### 2.3.1 Offline Debugging

Offline debugging techniques, also known as *post-mortem debugging*, typically analyse the execution of a program after it finished running. Offline debugging often involves capturing some contextual information of the program execution in a log for later analysis. Due to its simplicity, this approach is widely used in many modern applications, from cloud computing to operating systems. However, it is the responsibility of the developer to wisely choose what to log: capturing too little information may require many debugging cycles to find the root cause of the bug, while too much information may add too much noise to the analysis. In general, it is difficult to understand production failures from logs since extracting the right amount of relevant information about the failure often is very difficult[30].

Alternatively, *record and replay* debuggers trace the execution of a program and allow the developer to replay it afterwards.

Like log-based solutions, their scalability depends on the granularity of the trace [25]. Trace recording is especially challenging in concurrent and distributed systems due to non-determinism. Several replay debuggers exist to overcome these issues with different tradeoffs [14, 21, 28, 34, 36]. In order to deal with non-deterministic inputs, a partial order of variable accesses or events has to be stored in the trace to be able to reproduce the program concurrent behaviour. To make replay debugging scalable, several debuggers combine trace recording with *checkpoint-based debugging*, in which snapshots of the application are taken at certain periods of time to limit the size of the trace that needs to be stored [1, 16].

Reproducing a bug in our debugging scenarios with offline techniques can be tedious and time-consuming since many debugging cycles are required before the bug manifests again. In the twitter debugging scenario, it may not be viable to replay the application for hours to discover a bug. In the case of the CPS debugging scenario, the application works on continuous streams of data which may not be stopped. As such, it may not be convenient to let a temperature monitoring application crash, and retrieve a log or a trace of the execution afterward. In some cases, stopping the execution of the program may also lead to *information loss* [16].

### 2.3.2 Online Debugging

Online debuggers, often called *breakpoint-based debuggers*, allow users to control the execution of the target program and provide different facilities like pausing/resuming execution and step-by-step execution. To help investigate the root cause of a bug they allow developers to mark "interesting points" of the execution at which the program should pause, known as breakpoints. Once the program is paused, the debugger typically offers commands to (1) inspect the state of the program, often giving access to some data (e.g., a stack trace) that helps to understand how the program reached the current breakpoint, and (2) to control the debugged applications step-by-step by means of stepping operations such as stepping into or over a particular call.

In contrast to offline techniques, online debuggers can capture the context of a bug at the moment that it manifests. This makes them a good fit for our debugging scenarios since it allows the developer to analyze the state of her application without needing to replay it. However, object inspections and expression evaluations happen in the context of the running program, so if those expressions produce side-effects, those will be present in the program when the execution is resumed.

From an architectural point of view, we categorize online debuggers into two families: in-place (or in-process) and remote debuggers. Figure 1 shows the two architectures.

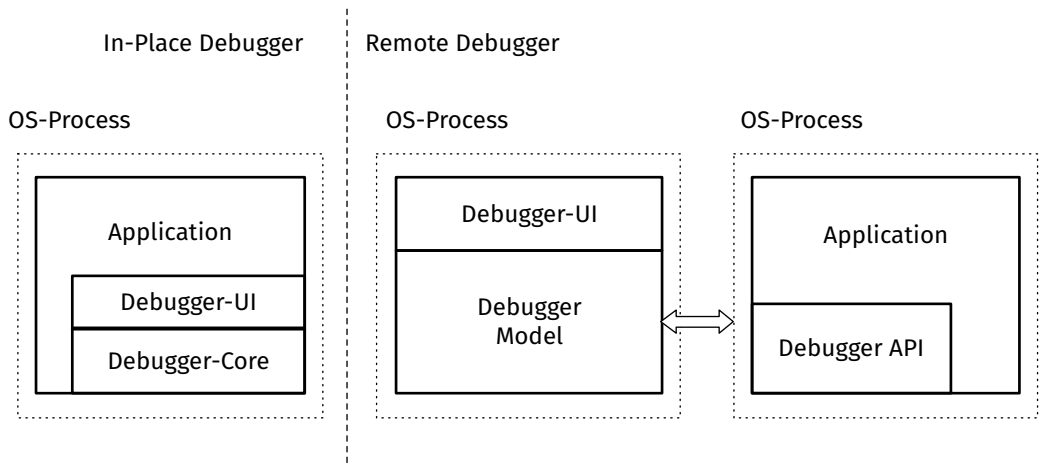

■ **Figure 1** In-place and remote debugging architectures.

**In-place Debuggers** An in-place debugger is an online debugger that executes in the same process as the application. It shares an address space with the application and can directly access its data and control its execution. As a result, developers can typically modify all objects of the application, including classes, instances, environments and in some cases runtime contexts. Examples of in-place debuggers include mainstream debuggers such as Python's, Perl's and Pharo's debuggers [2, 11, 33].

Since the debugger runs on the same process as the executed application, the developer does not experience latencies during debugging operations. This results in a generally good user experience since the debugger is highly interactive and provides immediate answers to the issued debugging commands.

On the other hand, to operate such debuggers, developers need to have direct access to the application process. For instance, connecting to Pharo's in-place debugger requires a screen and keyboard plugged into the machine being debugged. To overcome the need for direct access, a second architecture was designed: remote debuggers.

**Remote Debuggers** A remote debugger is an online debugger that controls the execution of the debugged application from a separate process (which we call the debugger process). The debugger process offers the same commands and features to the developers as an in-place debugger through its UI. However, since the debugger runs in a different address space than the debugged application, the debugger core is now split into the debugger-model at the debugger process and a debugger API at the application process (which controls the application according to the commands issued at the debugger process). Examples of such debuggers are JPDA [29] for Java,

GDB [15] for C/C++/Objective C, Visual Studio remote debugger [27] for .NET and TelePharo [20] for Pharo Smalltalk.

The main benefit of this architecture is that it allows the debugger to be deployed either on the same machine (typically a development scenario) or on a remote setup, deploying the two processes on different machines connected over a network. This makes them really useful for debugging scenarios like the CPS application in which access to the monitored machine is limited. However, all debugging operations in a remote debugger require inter-process communication between the debugger and the application process. As such, users may experience extra latency of the debugging operations for communication delays which rely entirely on the network performance and failures.

### 2.4 Problem Statement

As previously explained, offline debugging techniques can be tedious when logs are used, or they introduce a significant overhead in the case of record and replay [16]. More importantly, they require many debugging cycles to find the root cause of a bug since contextual information may not be present in the log or the recorded trace. On the other hand, online debuggers allow developers to directly debug a program while it is running, avoiding the tedium and additional overhead of execution-replay. They also allow developers to better understand the state of the application when a bug manifests since they capture the context of the bug and provide tools to further explore the program execution such as stepping commands, expression evaluation and state inspection.

The main drawback of online debugging solutions, however, is that they introduce *debugging interference* that may alter the behaviour of an application and affect the reproduction of a bug in a distributed and/or concurrent setting. We further distinguish the following kinds of interference:

**Latency** Breakpoints and stepping operations insert delays into the application execution to allow the developer to inspect the program state and understand it. Such pauses may affect the behaviour of concurrent programs and programs that have time-based constraints. These delays are higher when a debugger is deployed remotely, because it is connected to the program being debugged using network sockets or some other inter-process communication technique. This communication further increases the latency of the debugging operations

**Residual side-effects** During a debugging session, executing and inspecting arbitrary expressions may introduce side effects in the debugged program (e.g., assigning a variable, writing to an output stream). In traditional online debuggers, once such side-effects are applied, they are not rolled-back automatically and may affect the behaviour of the debugged program when it is resumed. Residual side-effects are problematic because they alter the application context, making it more difficult to reproduce the original bug. We say that side effects are *global* in online debuggers because they directly affect the application context

■ **Table 1**  Overview of debugging techniques based on the debugging interference, and their ability to capture bug context and operate remotely.

| Debugging Technique | Capture Bug Context | Remote Access | Latency | Side Effects | Examples |
|---|---|---|---|---|---|
| In-place | ✓ | ✗ | Low | Global | Pharo, Perl, Python |
| Remote | ✓ | ✓ | High | Global | JPDA,GDB |
| *Ours: Out-of-place* | ✓ | ✓ | *Low* | *Scoped* | *IDRA* |

Table 1 summarizes the different debugging techniques with respect to the debugging interference they present. On the one hand, in-place debuggers present a really low latency, because all operations happen locally, but do not provide remote access. On the other hand, remote debuggers provide remote access at the cost of degrading the user experience since all operations require inter-process communication. In this paper we aim to answer the following research question:

**Research Question:**  What online debugging architecture can (1) present the same latency as in-place debugging while (2) allowing remote debugging and (3) limiting the residual side effects?

To answer this question, we propose *out-of-place* debugging, a novel online debugging architecture. The main characteristic of out-of-place debugging is that it allows developers to debug an application remotely in isolation in a separate debugger process. As a result, debugging operations feature low latency and the residual side effects produced by the operations are limited to the debugger process. We say that side effects are thus *scoped* in out-of-place debugging. Section 3 further details out-of-place debugging and section 4 presents IDRA, an implementation of out-of-place debugging for Pharo Smalltalk.

## 3  Out-of-place Debugging

The main goal of *out-of-place* debugging is to debug remote applications with low latency and without influencing their execution. Similar to remote online debugging, an out-of-place debugger hosts the debugged application and the debugger in different processes. As such, the debugged application includes a debugging API in its infrastructure. However, in contrast to traditional remote debugging, out-of-place debugging transfers the entire debugging session (i.e. the application state and its state of execution) to the debugger process when the application reaches a breakpoint (or throws an unhandled exception). Once the developer finishes debugging, he can send a patch with the corresponding fixes and/or changes. The patch is dynamically applied to the running application and then the application is resumed. Developers can also discard a debugging session at any time without applying the code changes to the debugged application, and restart a new debugging cycle with the same exact original state.

In this section, we discuss the main components of an out-of-place debugging architecture.

### 3.1 Architecture Overview

Figure 2 shows the architecture of an out-of-place debugger. Like remote debuggers, an out-of-place debugger has components both on the debugger process and on the application process. In the application process, the out-of-place debugger infrastructure consists of the *Debugger Monitor* and the *Updater*. The debugger process, besides the debugger UI, is composed of the *Debugger Manager*, and the *Changes Handler*. In what follows we detail the role of each component when debugging an application. The numbers represent the order in which debugging operations take place when a breakpoint or exception halts the program's execution.

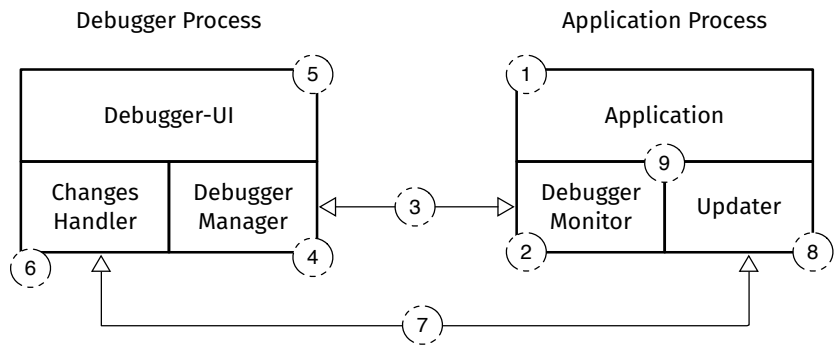

■ **Figure 2** Overview of an out-of-place architecture setup in two different processes. The arrow represents an inter-process communication channel.

**Debugger Monitor** The monitor implements the debugger API and is in charge of communicating with the debugger process. Its main roles are to supervise and control the application execution. When the application hits a breakpoint or raises an exception (1), the debugger monitor suspends the program execution. Then it extracts the state of the execution and the application's state, and creates a debugging session (2). The debugging session is then serialized and transferred to the Debugger Manager (3)

**Debugger Manager** The manager deserializes the application execution sent by the *Debugger Monitor* to recreate the debugging session and to generate the corresponding UIs that allow the user to debug (4). From the user perspective, debugging the application works similarly to an in-place debugger: through the Debugger-UI the developer can issue step commands, evaluate expressions and inspect the program state (5)

**Changes Handler** The role of the changes handler is to record all code changes done by the developer while interactively debugging the application (6).

Once the debugging cycle is finished, a commit operation is issued and the changes handler sends a patch with all recorded changes to the *Updater* (7).

**Updater**  The main role of the updater is to apply all code changes to the application that were recorded during the debugger cycle (8). It then notifies the debugger monitor which may resume the application execution (9) after updating the code

Note that the architecture of out-of-place debugging can be applied to debug both sequential and distributed applications. When an out-of-place debugger is instantiated for a sequential application, the architecture consists of two processes as depicted in fig. 2. When an out-of-place debugger is instantiated for a distributed application, the debugger consists of a debugger process and several application processes. The debugger manager can then be connected to more than one debugger monitor, handling different debugging sessions from different connected remote processes. This enables remote distributed debugging in a centralized way. This is the strategy used in IDRA, our out-of-place debugger described in section 4.

### 3.2  Creation of the Debugging Session

Out-of-place debugging aims to reduce the debugging interference by transferring all of the debugging operations to a different process than that of the debugged application. Transferring and re-constructing the debugging session on a different machine is crucial to the idea of out-of-place debugging. This enables an out-of-place debugger to (1) reduce latency during debugging, because all of the operations happen locally, thus avoiding network communication (2) scope the side effects of the debugging session to the process of the debugger.

In order to transfer the debugging session to a different process, the application state and its execution context need to be accessed and copied. The process of creating a copy of the debugging session at the debugger process is akin to *remote cloning* in the domain of code mobility [12]. We now further detail what creating a debugging session entails in the context of an object-oriented language.

In most object-oriented programming languages the application state is encoded as objects stored in memory, usually in the heap. The execution state is stored in a stack data structure, called the call-stack. Such a call-stack can reference objects in the heap. Figure 3 illustrates how the stack and the heap are related in the context of the Twitter debugging scenario. Each of the stack frames contains local variables that reference different objects in the heap. In some cases, the stack frame also points to the object that the method is executed on (receiver). The figure shows that the `analyseTweets` stack frame points to an instance of the `TwitterApplication` (receiver) and `Tweet` classes (local variable).

In order to create a debugging session, an out-of-place debugger walks the stack of the debugged application and selects all objects to serialize. Such selection is transitive, i.e. selecting all heap objects directly referenced from the stack implies selecting all objects referenced by these recursively. Moreover, the debugging session needs to also store all frame information that relates to the execution such as the name of the executed method and the program counter.

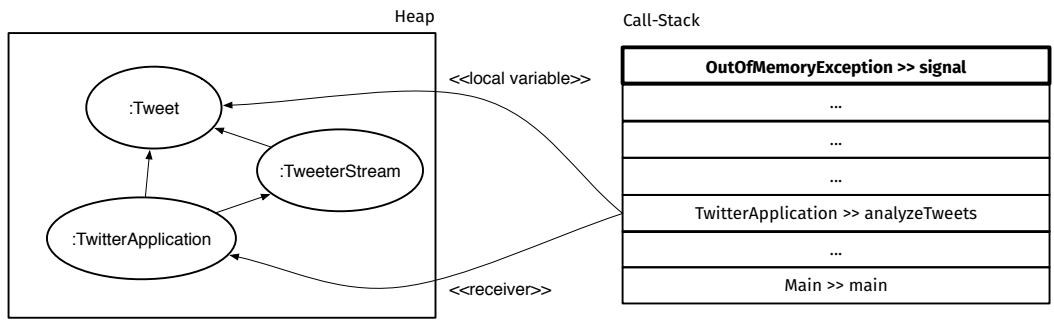

**Figure 3** Relation between the heap and the call-stack in the Twitter debugging scenario.

Similar to traditional remote debugging, out-of-place debugging assumes that the debugger environment has the same version of the code as the debugged application. This means that out-of-place debugger does not need to copy and serialize the executed code (i.e. classes or bytecode). Not only does this simplify the creation of a debugging session but it also reduces the amount of data transfers between application and debugger process.[4]

## 3.3 Allowing Scoped Remote Debugging

Out-of-place debugging allows for debugging an application remotely, keeping the benefits of a local debugging session. However, moving the debugging session completely onto a different process, while still enabling the basic properties of online debugging, poses two challenges: (1) how do we update the code of the remote application and (2) how do we access resources that are local to the remote application from the debugger process. To solve these challenges, we apply techniques from the fields of dynamic software updating and code mobility to out-of-place debugging. This is described in the remainder of this section.

### 3.3.1 Synchronizing the Code Base

After safely debugging their application in their local environment (i,e, the debugging process), developers should be able to send a fix to the debugged application. To this end, the changes handler of an out-of-place debugger records IDE interactions. It needs to keep track of code changes such as class and method additions, removals and modifications. These changes are stored sequentially in the order in which they were performed. Once developers finished applying changes to the code, they can issue a *commit*.

A commit operation packages all recorded changes in a single patch and sends it to the remote updater for its application. At the application process, the updater receives the patch and applies all the code changes that it contains in the correct order. The underlying platform should then be able to apply modifications to class structures, methods and objects in the running application. In this paper, we focus on the debugging aspects and assume that the underlying platform employs a mechanism for dynamically update software (DSU) to safely apply changes [18].

---

[4] In other words, an out-of-place debugger does not need to implement progress migration, which requires one to copy and transfer both code and execution state [12].

### 3.3.2 Handling Remote Resources

Out-of-place debugging offers in-place debugging on a remote debugger process. This raises the question of how to copy and transmit *non-transferable resources* used in the application code. These typically include objects representing external resources like files, sockets and sensors streams.

To exemplify the issues of handling remote resources, let us re-consider the Twitter debugging scenario. Listing 2 shows how the application reads tweets from a file: it checks a small header and reacts accordingly.

■ **Listing 2** Debugging a method accessing external resources.

```
1 TwitterApplication >> analyzeFileNamed: aName
2   | aFileStream header |
3   ...
4   aFileStream := (File named: aName) openForRead.
5   header := aFileStream next: 2.
6   (header == #(0 1) asByteArray)
7     ifTrue: [↑ aFileStream next: 10].
8   ↑aFileStream upToEnd.
```

Two different problems may arise if handling remote resources is approached naively:

1. If the debugger session is created before line 4 is executed, it will capture the execution before the creation of the file object, so no file object would be transferred in the debug session. However, stepping through line 4 in the debugger process will incorrectly open a file in the debugger process instead of the application's process

2. If the debugger session is created after line 4 is executed, it will capture the execution after the creation of the file object. A file stream will be copied and transferred to the debugger process. The transferred file stream will have a reference to a file descriptor that is invalid in the developer's machine where the debugger process runs. Again, the user will get a failure when a step operation tries to access the file stream locally

In order to avoid manually instrumenting the code that uses non-transferable resources, our out-of-place debugger proposes the use of remote proxies to allow access to remote resources from the debugger process[13]. This approach is similar to the use of network references in the domain of code mobility as described by Fuggetta, Picco and Vigna [12]. Remote proxies can be transparently introduced during the creation of the debugging session using two techniques: serialization-time object substitution and code instrumentation.

**Serialization-time Object Substitution**  Upon serialization, a substitution rule needs to be provided to the object serializer to replace an external resource object by a cor-responding remote proxy to it. Such a proxy can then be accessed in the reconstructed debugging session.

**Code Instrumentation**    To avoid local allocation of non-available external resources, all accesses to remote resources should be captured. Code instrumentation techniques allow us to substitute all accesses to pre-defined classes (such as *File*) with the instantiation of a proxy. This operation can happen transparently to the developer and is not visible in the code. Upon applying instrumentation, the developer is able to access the original file through the proxy.

Section 4.3 shows how our out-of-place debugger implements these mechanisms for file streams, and optimizes them with buffered reads.

## 4    IDRA: a Prototype Implementation of *out-of-place* Debugging

We implemented IDRA[23], a concrete instantiation of out-of-place debugging for the Pharo Smalltalk [32] programming language. We choose Pharo as our experimentation platform because it features an extensible environment to quickly prototype our debugging architecture. More concretely, the Pharo platform reifies the stack of execution as Pharo objects, and provides a debugger written in Pharo itself which is accessible and can be extended.

We strongly believe that the *out-of-place* debugging architecture can be implemented for other programming platforms that (1) provide reflective capabilities that reify the execution stack and provide access to a debugging interface, or (2) allow to introduce virtual machine modifications (when reflective capabilities are limited).

### 4.1  Architecture of IDRA

The architecture of IDRA has the same building blocks as the ones present in out-of-place debugging explained in fig. 2. The main three IDRA components are:

1. the *IDRA Monitor*, an instance of the Debugger Monitor at the application process,
2. the *IDRA Manager*, an instance of the Debugger Manager at the debugging process, and
3. the *IDRA Changes Handler*, an instance of both Changes Handler and Updater

The IDRA Manager allows debugging different sessions coming from different connected IDRA Monitor(s). The communication between the different instances happens either via message passing on TCP sockets, where the connection needs to be kept alive, or via REST HTTP calls through the Zinc HTTP library [37]. All data serialization happens through the object serialization library Fuel [8], which is able to serialize standard user-created objects as well as meta-objects such as the call-stack.

The IDRA Changes Handler is also deployed on both application and debugger processes, and is capable to exchange code changes between two (or more) instances. Communication between the two IDRA Changes Handlers also happens by means of the two aforementioned networking technologies.

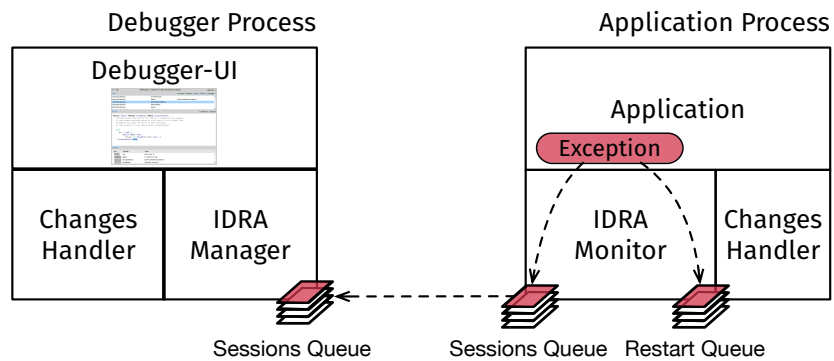

■ **Figure 4**   Architecture of IDRA exception handling.

The IDRA Changes Handler and the two entities of the debugger (IDRA Manager and Monitor) are, voluntarily, separated. For instance, they use two different communication channels. This allows for deploying the changes handling infrastructure separated from the debugger, and enables to use it in other use cases, such as simply updating the code of a remote image.

### 4.2 Handling of Exceptions and Code Changes

Inspired by the Smalltalk tradition, IDRA models both breakpoints and unhandled exception as exceptions which suspend (*halt* in Smalltalk terminology) the program execution. In order to potentially handle, on the application side, concurrent exceptions and, on the debugger side, debugging sessions from different monitors, both IDRA Manager and Monitor handle exceptions asynchronously. To this end, IDRA Manager and IDRA Monitor share the same basic mechanism: they feature a queue, called *sessions queue*, where debugging sessions are stored. The IDRA Monitor queue actually stores the debugging session created upon an exception which is then later sent to the IDRA Manager. On the other hand, the IDRA Manager pops debugging sessions from its queue to open a debugger to the user.

Figure 4 shows in detail how IDRA handles the exceptions when it suspends the program execution. The IDRA Monitor stores the debugging session in two different queues: the *session queue*, so the outgoing queue, and the *restart queue*. The restart queue is needed to keep track of the debugging sessions that were already sent to the IDRA Manager. It allows IDRA to restart the failed debugging session after the developer commits its fix to the remote machine. Figure 5 shows how, after the developer produces and commits a fix, this triggers a re-execution in the debugged application. This re-execution phase can happen following different strategies, depending on the use case. For instance we provide specific restarting strategies for test execution, or for tasks scheduled using the TaskIt library [3].

To correctly detect code changes in the debugger project, the IDRA Changes Handler leverages on Epicea [7], an existent library for handling such events.

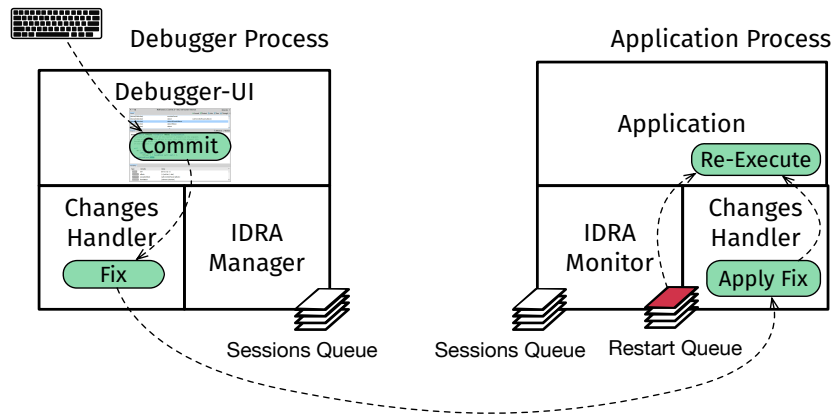

**Figure 5** Architecture of IDRA code changes handling.

### 4.3 Handling Remote Resources

We now detail how we implemented the two strategies for handling remote resources described in section 3.3.2: object substitution during serialization (when we are transferring an already referenced remote resource) and code instrumentation (when a remote resource needs to be opened remotely). We implemented in our prototype a proof-of-concept to handle remote files. Our current implementation does support read streams but not yet write streams, nor other external resources such as socket streams or sensors; this is ongoing work.

**Object substitution**    During the serialization of a debugging session, while navigating through the stack, we substitute all the instances of `FileStream` with a corresponding `RemoteFileStream` object, which will contain information on how to set up a proxy. When the object is reconstructed, the proxy will be automatically set up. A table of remote resources is kept at the application side to keep track of the different proxies.

**Code instrumentation**    We implemented code instrumentation using meta-links [6], specialized meta-objects that control the execution of AST nodes. During execution, a metalink can execute its code before, instead or after the execution of its annotated AST node. This facility gives fine-grained instrumentation at the sub-method level. We use metalinks to transparently replace all accesses to external resource classes by accesses to a corresponding `Remote*` proxy class. In our concrete implementation for files, we replace all accesses to the `File` class by calls to `RemoteFile`. On the first access, such `RemoteFile` will retrieve from the `IDRA Manager` the necessary information to setup a proxy with the remote application. Then during debugging, the `RemoteFile` proxy class manages the access to the remote file.

To optimize our solution for files, we implemented `RemoteFileStream` as a buffered proxy that minimizes network roundtrips. When reading out of the bounds of the buffer, new contents for the buffer are retrieved, allowing the developer to keep reading on the original file. Our approach can handle remote files without the need of transferring it entirely, still providing direct access to it through a buffered communication.

## 5 Validation

In this section we validate out-of-place debugging by evaluating IDRA with respect to performance and usability. In particular, we compare IDRA to TelePharo [20], the remote debugger for Pharo programs [31]. TelePharo can be seen as state-of-practice mainstream technology in remote debugging. It uses proxies on the debugging session (and everything that it references) to allow debugging from a different process.

We compare both approaches by means of the debugging scenarios described in section 2. We evaluate both IDRA and TelePharo debugging features on those scenarios *quantitively*, using performance benchmarks on initialization time and communication overheads, and *qualitatively*, analysing the debugging experience in the CPS scenario from section 2.2.

### 5.1 Quantitative evaluation

The goal of this section is to evaluate the performance of IDRA. Appendix A details the benchmark conducted to evaluate the debugging latency. Since IDRA behaves as an in-place debugger while developer debugs the application, the latency on different debugging operations in IDRA is naturally lower than the one of a remote debugger such as TelePharo. As such, the focus of this section is to evaluate IDRA in terms of communication and performance overhead. In particular, we focus on evaluating the impact of copying and transferring a debugging session in an out-of-place debugger versus employing a traditional remote debugger. To this end, we compare both debuggers in terms of:

1. The time to initialize a debugging session
2. The impact of remote debugging in terms of data transferred between the debugger and application process
3. The impact of propagating code changes in terms of data transferred between the debugger and application process
4. The performance overhead of IDRA on the target application system

**Setup** The benchmarks were executed on two different machines: the *debugger machine*, the machine where the debugger is deployed, and the *worker machine*, the machine where the application is deployed. Both machines are a four-core Intel® Core™ i7 Processor at 3.5 GHz with turboboost, and 16 GB RAM DDR3. Both machines run Apple macOS (version 10.13.4) with Pharo 7.0+alpha (Build 848), 32 bit. They are connected through a WiFi 2.4 GHz local network. All benchmarks that compare IDRA with TelePharo employ the Twitter analyzing application described in section 2.1 where we read tweets from a file, parse them and create an instance.

■ **Table 2** Session initialization time (in milliseconds).

| Debugger | $T_{queue}$ | $T_{materialize}$ | $T_{replay}$ | $T_{init}$ |
|---|---|---|---|---|
| TelePharo | - | - | - | *0.14* |
| IDRA | 62.88 | 1.03 | 0.41 | *64.32* |

### 5.1.1 Session initialization time

This benchmark assesses the impact of out-of-place debugging on the set-up of a single debug session. It measures how long it takes for each debugger to open a debug session (i.e. the time between the VM receives an exception and a debugger UI is opened). We do not consider the time needed to render the debugger UI.

TelePharo opens immediately a debugger UI for each debug session it receives, while IDRA opens only one debugging session at a time. To have an equivalent evaluation, we closed a debugging session opened by IDRA before opening another one.

**Results** Table 2 shows that, on average, TelePharo initializes a debugging session in the order of nanoseconds, while IDRA does it in the order of milliseconds. This difference is normal because TelePharo does not perform any operation other than requesting an UI for the debugger (and the UI rendering time is not considered).

Table 2 breaks down the initialization time of IDRA in the time to (1) enqueuing the debugging session , (2) materializing the serialized debugging session, and (3) finally re-playing it. We observe that IDRA actually spends most of its initialization time due to the queuing of debugging sessions. As explained in section 4, in order to support multiple debugging sessions, IDRA queues debugging sessions and opens only one at a time. Debugging sessions are extracted asynchronously one by one from the session queue, with an average waiting time of at least 60 ms.

The results of this benchmark also show that IDRA provides the developer with the opportunity to completely replay from scratch a debugging session in a negligible amount of time (~0.41 ms on average), since the debugging session is already serialized and stored. In contrast, TelePharo requires the developer to manually start a new debugging session and reproduce the bug before getting to the same previous state.

### 5.1.2 Overhead of remote debugging

In this section we compare the overhead of IDRA and TelePharo when transferring debugging sessions over the network. To do so, we measure the amount of network communication employed to transfer the necessary information when one exception is raised at the application process between the *monitor-manager* in the case of IDRA, and *server-client* debugger infrastructure in the case of TelePharo. Note that IDRA transfers all the information of an exception when this is handled at the application process. On the other hand, TelePharo first transfers proxies to control the application,

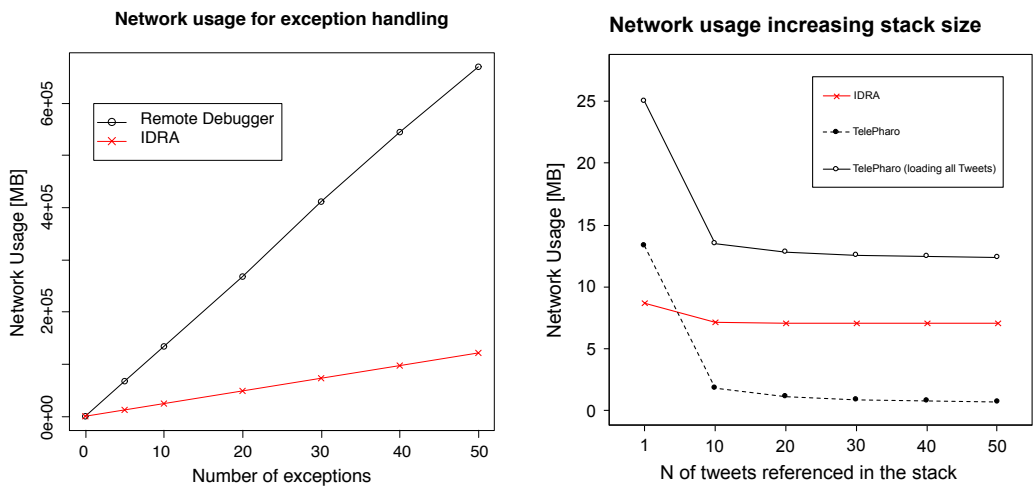

**(a)** Number of bytes exchanged for an increasing number of debugging sessions.

**(b)** Number of bytes exchanged for an increasing the size of the stack

■ **Figure 6** Result of the benchmarks for network overhead.

and only transfers data through them as needed by the different debugging operations. This benchmark measures the amount of bytes exchanged[5]:

(a) when sending an increasing number of exceptions with a fixed stack size

(b) when sending a fixed number of exceptions, with an increasing stack size

We employ the twitter application in which an exception is thrown after the worker parsed the JSON string of a tweet. At the moment the exception is thrown, the stack references both the string(s) related to the tweet(s) it is parsing, and the references to the tweet object(s) that have been instantiated. For benchmark (a), only one tweet is parsed (and referenced). For benchmark (b), we increase the number of tweets parsed by a single worker, from 1 to 50, meaning that the stack will reference 1 to 50 tweet strings and objects. Benchmark (b) always analyzes a total of 600 tweets, constant in the different iterations.

**Results** Figure 6 shows the amount of bytes exchanged for benchmark (a) and (b). Figure 6a shows that, on a small sized stack, both debuggers overhead grow linearly with the amount of data sent. However, IDRA performs better because the application stack and the associated data are smaller than all the communication necessary to install and exchange proxies.

The results of benchmark (b) in fig. 6b give more details on how the stack size influences the amount of data transferred on the two debuggers. The X-axis represents how many tweets are referenced by the debugged stack. Since tweets are grouped in the different executions, X=1 will mean 600 debugging session opened, X=10 will open 60 debugging session, etc. The value on the Y-axis is the total number of bytes for processing all the 600 tweets. The results show that the amount of data transferred (summing up all the debugging sessions) is more or less constant around 8 MB, although it decreases slightly when increasing the number of tweets per session. This means that, as expected, the amount of data transferred is related to the amount of tweets referenced by the stack, with a low communication overhead (in the order of kBs) when increasing the number of exceptions.

---

[5] We measure amount of bytes because the transferring time depends on too many variables such as speed of the network, congestion, failures rate, etc.

Figure 6b compares the number of bytes exchanges in TelePharo in two ways. The dashed black line shows the amount of data transferred by TelePharo when sending the exceptions (this entails sending proxies of the references objects, an opening a debugger on each exception). The full black line shows the amount of data transferred by TelePharo when sending the exceptions, but also accessing the proxied values of the instantiated tweet objects. This reflects the behaviour of TelePharo when a developer loops over the tweets and accessing their real value.

Both values are really high when there are 600 sessions with only one tweet, but decrease when increasing the size of the stack and decreasing the number of exceptions. This is because in TelePharo the amount of data transferred is not directly related to the number of objects referenced by the stack, but to the amount of debugging sessions. The difference between the dashed and full black lines shows the impact of retrieving real values of the tweet objects. The dashed line shows that sending proxies of the referenced objects does not impact much the amount of data transferred when opening the debugging session. The full black line shows an offset of a fixed value, that is the amount of communication necessary to actually retrieve the real value of the tweet objects.

From this benchmark we can conclude that the amount of communication overhead of the two debuggers is definitively comparable: IDRA's network usage is better than TelePharo's when debugging a small stack, and as expected, worse with bigger stack sizes. However, if during a TelePharo debugging session the developer inspects objects, the network overhead is higher than when transferring the full debugging session with IDRA, since accessing proxied objects requires more network communication. Hence, IDRA represents a good alternative to TelePharo even with big stacks, when developers access the different objects referenced in the debugging session.

### 5.1.3 Overhead of propagating code changes

To assess the performance of handling changes in out-of-place debugging, we compare the overhead of IDRA and TelePharo when transferring code changes. Both debuggers employ very different approaches to handle code changes. In IDRA, changes happen locally and are then sent to the remote machine through the IDRA Changes Handler in a single network transmission when the user explicitly performs a commit. In contrast, in TelePharo the user directly modifies via a remote browser (or a remote debugger) the classes of the remote machine, and changes are directly applied to the application process.

In this benchmark we compare the network overhead of triggering a code change in the remotely debugged application. For different code changes, we measure the bytes of propagating:

**No operation.** No changes are made. A browser is opened and changes are sent

**A class addition.**  A class named Test01 is added to a package
**An instance variable addition.**  An instance variable named instanceVariable is added to Test01
**A class variable addition.**  A class variable named classVariable is added to Test01
**A method code change.**  A method of the class Test01 is changed adding a line of code

**Results**    Table 3 shows the amount of bytes (in KB) used to remotely apply single code changes. The network overhead when not changing code is approximately the same, showing no significant difference between both debuggers. In all other operations IDRA uses 8 to 10 times less network when compared to TelePharo for simple committing operations. Since TelePharo uses a remote browser (stand-alone or included in the remote debugger), every modification constantly generates a request to the remote image to update. This does not happen when using IDRA because the changes happen first in the local code base, and then are packaged and transferred to the remote image, resulting in less communication overhead.

■ **Table 3**    Network usage for committing single code changes

| Operation | IDRA [KB] | TelePharo [KB] |
|---|---|---|
| No operation | 0.9 | 1.1 |
| Class addition | 2.5 | 20.4 |
| Instance Variable addition | 3.3 | 24.1 |
| Class Variable addition | 3.6 | 31.4 |
| Method Change | 2.5 | 41.2 |

#### 5.1.4  Overhead of IDRA on debugged applications.

In order to quantify the impact of IDRA on a running application, we compare the execution with and without an active IDRA instance in the worker machine. To this end, we run the Computer Language Benchmark Game [35] in the implementation available for SMark [22], a benchmarking framework for Pharo. Such benchmarks measure the performance of the overall system applying common programming problems such as n-body, K-nucleotide and thread-ring. We measure the overhead of both the IDRA Monitor, and the IDRA Changes Handler.

**Results**    We executed each of these benchmarks one hundred times in two different set-ups: (a) in a worker machine where IDRA was not installed and (b) in a worker machine with IDRA initialized and connected to the IDRA Manager at the debugger process. The averages were calculated for each of the benchmarks, and the ratio was calculated dividing the two averages.

Table 4 shows that IDRA does not introduce significant overhead in the execution. Although our benchmarks represent a hundred executions, on average some benchmarks are equal, in some the baseline is faster and in some the benchmark is even faster when IDRA is running. This makes us think that differences in the execution time are not related to the presence of the debugger.

■ **Table 4** Overhead of the execution with active IDRA Monitor.

| Benchmark | Baseline[ms] | IDRA Active[ms] | Ratio |
|---|---|---|---|
| SpectralNorm | 10,44 ±0,21 | 11,20 ±0,16 | 1,07 x |
| ThreadRing | 0,56 ±0,16 | 0,54 ±0,16 | 0,96 x |
| Nbody | 1,48 ±0,18 | 1,69 ±0,19 | 1,14 x |
| Mandelbrot | 32,95 ±0,33 | 39,52 ±0,21 | 1,20 x |
| ChameneosRedux | 29,40 ±1,90 | 22,44 ±0,23 | 0,76 x |
| RegexDNA | 405,40 ±4,00 | 414,37 ±0,61 | 1,02 x |
| Meteor | 214,80 ±3,70 | 213,20 ±2,10 | 0,99 x |
| Fasta | 1,15 ±0,19 | 0,83 ±0,18 | 0,72 x |
| BinaryTrees | 3,49 ±0,29 | 2,51 ±0,16 | 0,72 x |
| Chameleons | 3,60 ±0,21 | 3,74 ±0,25 | 1,04 x |
| ReverseComplement | 0,45 ±0,15 | 0,43 ±0,15 | 0,96 x |
| Knucleotide | 17,76 ±0,34 | 15,88 ±0,18 | 0,89 x |

## 5.2 Qualitative evaluation: Assessing the CPS debugging scenario

In this section, we compare the debugging experience in IDRA and TelePharo in the context of the CPS scenario described in section 2.2. This debugging scenario shows the challenges when debugging a modern application which analyses a continuous stream of data and is running on a resource-constrained device which may not be accessible to the developer. The scenario steams from an unrelated research project whose goal is to program and debug CPS applications in Pharo conducted by Costiou [5]. During the presentation of that work, the main developer reported problems when debugging the temperature monitoring application deployed remotely on a raspberry-pi with TelePharo. We then collaborated together in a case study to assess whether IDRA could help him to find the root cause of his bug.

This case study allows us to gain initial insights into how the key properties of out-of-place debugging (i.e. low debugging latency and scoped size effects) can help debugging modern applications. It remains future work to conduct a systematic qualitative evaluation by means of a user study to assess the usability of IDRA. In what follows we discuss the debugging experience of IDRA and TelePharo. Further details on the case study can be found in [24].

### 5.2.1 Debugging experience using TelePharo

The CPS developer first tried to use TelePharo. While he was debugging, he was experiencing two different problems, both symptoms of the high debugging interference:

**Big delays during debugging.** There were big delays while applying any debugging operation (stepping, or expression evaluation). This caused a lack of immediate feedback and increased the debugging time;

**Errors in the deployed application.** Multiple times, he introduced an error in the application code provoking a failure in either his local debugging instance, or in the remotely deployed application. This broke the debugging connection, hanging the remote environment and forcing him to manually reboot his remote system

Due to these problems the developer could not find the root cause of his bug.

### 5.2.2 Debugging experience using IDRA

With IDRA, the developer was able to debug his application locally at his machine starting from the original exception, experiencing low latency on the debugging operations. During the debugging he changed the code and tested it locally before deploying it back into the remote machine. Most importantly, he avoided crashing the application on both his local and remote machine. As a result, his debugging experience in comparison to TelePharo was improved and he found out the root cause of his bug: in some occasions the sensor was returning a string containing "nan" instead of a concrete value, meaning that the sensor read from the external driver failed. Such value was parsed as a number, causing a parsing exception.

Besides fixing the bug in the monitoring application, the developer also used IDRA to debug different applications running on his remote machine, from an unique centralized point. This was possible because IDRA enables developers (1) to handle different monitors from an unique manager, and (2) to handle multiple exceptions through a queuing system that allows them to selectively debug remote exceptions.

## 6   Related work

In this section we compare out-of-place debugging to other solutions which combine remote online and offline techniques. To the best of our knowledge, the closest related work is in the fields of web-applications and Big Data. In particular, we review replay debuggers that allow to debug remotely like out-of-place debugging, and hybrid approaches, based on replay/checkpointing, that still feature some online debugging operations such as breakpoints.

Jardis [1] is a post-mortem debugger for JavaScript/Node.JS that uses checkpointing, taking snapshot every two seconds. In this way it can limit the size of the trace discarding all the previous ones, hence allowing to replay only for the last 2 seconds. Interestingly, it is deployed in a similar architecture than out-of-place debugging, allowing to debug a remote cluster. However, it transfers information using an intermediate log, and allows developers then to debug in a fresh post-mortem environment.

Daphne [19] is a debugger for DryadLINQ [26]. It provides a runtime view of the system and the query nodes generated by a LINQ query. It allows developers to add breakpoints to inspect the state and start and stop commands through the Visual Studio remote debugger. Debugging is done directly on the client where the breakpointed node is executing, interrupting it in order to debug it. Execution of a specific node can be replayed locally to analyze its execution using the same debugging primitives.

BigDebug [16] is a debugger for Apache Spark [10] which introduces the concept of *simulated breakpoint* that does not stop the execution nor freezes the system waiting for the resolution of the breakpoint. Instead, it stores the information necessary to replay the environment and then continues the execution. BigDebug also provides *watchpoints* using *guards*. A watchpoint monitors some expression using a predicate function. The developer can then lively visualize the value of the watched expression when the predicate is satisfied.

While these debuggers help the developer in finding a bug in scenarios similar to the described in this paper, they still fail to correctly capture the context that produced a bug. Anyhow, a replaying step (at least) from the latest checkpoint is needed. When online capabilities are allowed (i.e. simulated breakpoints in BigDebug), they are still employed after a replaying step.

## 7   Conclusion

In this paper we presented *out-of-place* debugging, a novel online debugging architecture that allows developers to debug remotely, keeping the debugging experience of in-place debugging. It does so by transferring the entire state of the program execution (application objects and call-stack) to a different process executing on the developer's machine. Since the debugging operations apply locally, our approach minimizes the latency created by network communication and scopes all the side effects to the local debugging process.

We prototyped our approach implementing IDRA, an out-of-place debugger for the Pharo Smalltalk programming platform. Our implementation handles external resources by the usage of object replacement during serialization and code instrumentation using metalinks. We showed that *out-of-place* debugging is practical by performing both quantitative and qualitative evaluation, comparing IDRA to Pharo's remote debugger in two different debugging scenarios. The benchmarks show that IDRA's remote capabilities perform well when compared to TelePharo, and that the presence of our debugger does not impact the overall performance of an application.

For future work, we plan to investigate how this architecture can be applied in the Big Data setting, handling highly parallel applications that analyze continuous streams of data. We also plan to release IDRA to the Pharo community, allowing us to test it in realistic industrial scenarios.

**Acknowledgements**   We would like to thank the anonymous reviewers for their constructive comments. We would also like to thank Jennifer Sartor, Clement Bera, Carmen Torres Lopez and Pablo Tesone for their help reviewing early versions of the paper.

Matteo Marra is a PhD-SB fellow at the Fonds Wetenschappelijk Onderzoek - Vlaanderen - Project number: 1S63418N. We are also grateful to the financial support of the European Smalltalk User Group (http://www.esug.org).

## Appendix A

In this appendix we present an additional benchmark in which we compare the performance of single stepping operation between IDRA and TelePharo. This benchmark is included here for completion, to show how much IDRA reduces the debugging latency during debugging. It corresponds to a direct comparison between using an in-place approach and a remote one. This benchmark is executed in the same environment as the rest of the experiments described in section 5.1.

**Latency of stepping operations**
When a debugging session arrives to the debugger process, both IDRA and TelePharo open a debugger on it. At this point different operations can be executed:

**Restart.** Alters the call-stack to be at a selected stack frame, discarding newer frames

**Step Into.** Resumes the execution until the start of a new stack frame

**Step Over.** Resumes the execution until it returns to the same stack frame

**Step Through.** Resumes the execution until it returns to the same stack frame or enters a stack frame of a locally created closure

**Proceed.** Resumes the normal execution

We benchmarked each operation as follows:

1. *Restart* from a point in the stack
2. Execute the operation (*Step into/over/through*). This step is not executed in the case of the *Restart operation*
3. *Proceed* the computation

This actions represent a typical debugging session, except the fact that no code is changed. It is however consistent to evaluate the execution time of the operations on both debuggers.

**Results**    Figure 7 shows the results of our benchmark. The execution time is represented in a logarithmic scale. IDRA is consistently faster than TelePharo between a hundred and a thousand times. In fact, in the case of a debug session handled with IDRA, the exception and all the stack information is copied and sent to the debugger. A debugging session is always opened on a local copy of the exception (and its stack), which makes the debugging session a normal in-place Pharo debugging session.

On the other hand, TelePharo reconstructs a remote exception by means of proxies of the exception itself and of the related stack. The debugging operations will be executed on the remote machine, introducing communication and network overhead for each of the operations executed.

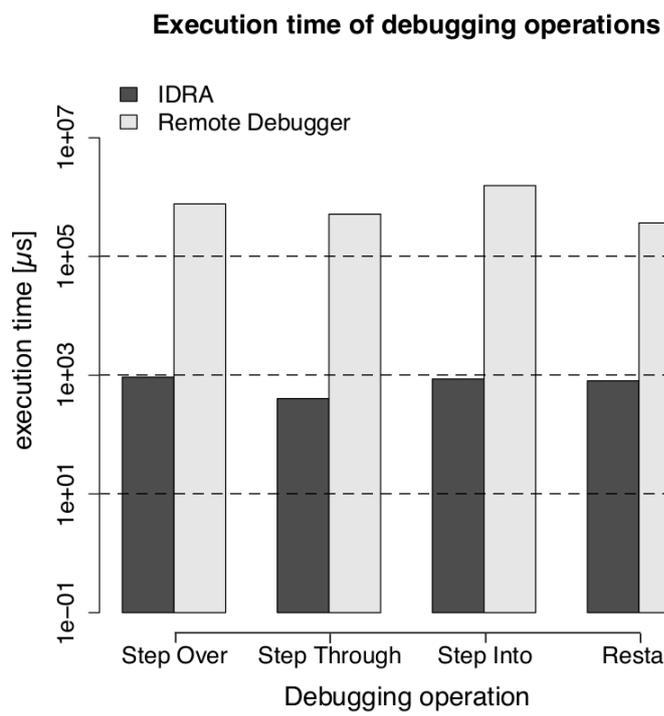

■ **Figure 7** Bar plot of the execution time of single debugging operations.

## About the authors

**Matteo Marra** is a PhD student at the Software Languages Lab of the Vrije Universiteit Brussel. He got his master in 2016 at the same university, and is currently conducting his research over debugging big data applications. This paper mainly represents the work of his master thesis. You can contact him at mmarra@vub.be.

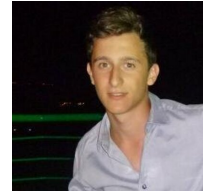

**Guillermo Polito** is a research engineer at the CRIStAL laboratory in the university of Lille, working in tight relation with the RMoD team. Guille's main research interests are programming language development, programming tools, modular systems, and maintenance of large software systems. He currently works on techniques to develop modular systems and languages, and new development tools. Guille participates in the development of the open source Pharo programming language and environment since 2010. You can contact him at guillermo.polito@inria.fr.

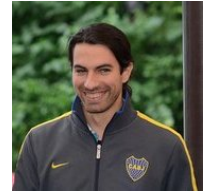

**Elisa Gonzalez Boix** is an Assistant Professor at the Software Languages Lab (SOFT) of the Vrije Universiteit Brussel (VUB), Belgium. She obtained her Master in Informatics Engineering in 2004 from the Universitat Politecnica de Catalunya (Spain) and her PhD in Sciences in 2012 from VUB on programming language abstractions and tools for handling partial failures in distributed applications running on mobile ad hoc networks. Her PhD heavily relied on reflection and meta-level programming. Since 2014, she leads a a group on concurrent and distributed systems, studying programming abstractions and dynamic software tools like debuggers. You can contact her at egonzale@vub.be.

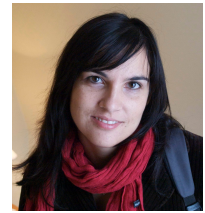